\documentclass[12pt]{iopart}

\usepackage{hyperref}
\usepackage[dvips]{graphicx}
 
\def\v#1{{\bf#1}}
\def\be{\begin{equation}}
\def\ee{\end{equation}}
\def\bea{\begin{eqnarray}}
\def\eea{\end{eqnarray}}
\def\ahalf{{\textstyle{1\over2}}}

\newcommand{\bfalpha}{\mbox{\boldmath$\alpha$\unboldmath}}

\newcommand{\bfsigma}{\mbox{\boldmath$\sigma$\unboldmath}}
\newcommand{\bfSigma}{\mbox{\boldmath$\Sigma$\unboldmath}}

\newcommand{\bfphi}{\mbox{\boldmath$\phi$\unboldmath}}
\def\ie{{\it i.e.\,}}

\def\fcal{\mbox{$\cal F\,$}}

\def\<{\langle}
\def\>{\rangle}

\begin{document}

\title[Dynamics of a Dirac oscillator coupled to an external field]{Dynamics of a Dirac oscillator coupled to an external field: A new class of solvable problems}

 
\author{E. Sadurn\'i$^1$, J.M. Torres$^1$ and T. H. Seligman$^{1,2}$}

\address{$^1$Instituto de Ciencias F\'isicas,
Universidad Nacional Aut\'onoma de M\'exico,
Cuernavaca, Morelos, M\'exico.}

\address{$^2$Centro Internacional de Ciencias,
 Cuernavaca, Morelos, M\'exico.}

\eads{ \mailto{sadurni@fis.unam.mx},\mailto{mau@fis.unam.mx},\mailto{seligman@fis.unam.mx}}

\begin{abstract}

The Dirac oscillator coupled to an external two-component field can retain its solvability, if couplings are appropriately chosen. 
This provides a new class of integrable systems. A simplified way of solution is given, by recasting the known solution of 
the Dirac oscillator into matrix form; there one notices, that a block-diagonal form arises in a Hamiltonian formulation.
The blocks are two-dimensional. Choosing couplings that do not affect the block structure, these blow up the $2 \times 2$ matrices to $4 \times 4 $ matrices, thus conserving solvability.
The result can be cast again in covariant form. By way of example we apply this exact solution to calculate the 
evolution of entanglement. 

\end{abstract}

\pacs{03.65.Pm, 03.67.Mn, 12.90.+b}
\submitto{\JPA}
 
\maketitle
 
\section{Introduction}
 
The Dirac oscillator \cite{DO} has attracted considerable attention both due to its simple 
formulation and its analytical solutions. Yet we may ask, if we can go beyond the single 
particle problem and conserve some of that elegance and simplicity. A plausible scenario 
in particle physics is that of a Dirac particle bound by an oscillator and interacting with an external (free)
non-abelian field of mesons modeled as particles with finite radius \cite{yukawa1,yukawa2}.
Specifically, if the internal group of the interaction is chosen to be $SU(2)$ and the non-local 
potential dependence is linear in the position and momentum operators, the structure which leads 
to integrability is preserved. This linearity can also be viewed as an approximation to a more complicated field. The central subject of this paper will thus be to develop this scenario for one, two and three dimensional Dirac oscillators. 

To achieve this we shall reformulate Moshinsky's approach in a matrix representation, where we shall see that a simple transformation 
will take the Hamilonian formulation of the Dirac oscillator to a  block-diagonal form with $2 \times 2$ blocks.
This representation is not manifestly covariant due to its Hamiltonian form. The equivalence which follows from the spectral 
theorem is confirmed by the fact that we recover the well-known solutions. The coupling to the field, which carries isospin, will necessarily lead to $ 4 \times 4$ blocks, which can still be solved exactly. The solvable models are obtained by choosing the coupling in such a way, that the blocks remain uncoupled.
We shall show that several interesting models can be derived in this way, and that the two-dimensional case has a special 
role as compared to the one and three-dimensional cases. A covariant formulation of the different cases is  also given. Once we obtain the exact eigen functions and eigenvalues we can study dynamics.

Several cases of interest are studied by choosing particular values for the parameters of the model. For instance, we can have a coupling to the field including explicit dependence on $\gamma$-matrices (Pauli coupling). The conservation of isospin + fermionic charge in the dynamics can be produced by restricting this example. The case of vanishing coupling with the external field except for its static part is included as the simplest extension of the Dirac oscillator. The absence of explicit fermionic rest mass in the hamiltonian is also interesting in the sense of mass generation due to the external field.

By way of example and as an extension to the 
fidelity studies presented for the Dirac oscillator in \cite{JPA2008} we take advantage of the additional degree of freedom and consider the evolution of entanglement between the oscillator and the field or, equivalently, the decoherence of the oscillator due to the field.

In section two we will review the Dirac oscilator in one, two and three dimensions in a matrix representation 
and show, that it is solved by casting it in block-diagonal form, the blocks being two-dimensional. This will not only serve as a reminder of previous work and fix our notation. It will also set the stage for the generalization given in the following section. There we will implement the coupling to the isospin field 
and show that this can be done such that the block size is increased only to four. This leaves us again with solvable models.
In section four we give the covariant formulation of these models and in section five an in-depth discussion of the one-dimensional case.
To get a feeling of the solvable dynamics we have found, we calculate in section six the evolution of entanglement 
between the field and the Dirac oscillator. We shall end by drawing some conclusions and giving an outlook.

\section{The algebraic structure of 1,2 and 3 dimensional Dirac oscillators}

The Dirac oscillator \cite{DO} was originally proposed as a way to introduce a linear potential \cite{ito} which preserved the solvability of the resulting Dirac equation. We review some of the known results of this system by using a new notation. The treatment to be presented is related to the algebraic structure of the Dirac oscillator hamiltonians in 1, 2 and 3 dimensions. The relation of this structure with the solutions of the associated eigenvalue problem is discussed. This should be helpful when we consider solvable extensions of this system. Let us start with the three dimensional case, originally considered by Moshinsky and Szczepaniak. The hamiltonian is

\bea
H= \bfalpha \cdot \left(\v p + i \beta \v r \right) + m \beta
\label{0.1}
\eea
where we have adopted units such that the rest mass and the frequency satisfy $m \omega=1$ and $\hbar=c=1$. We use a representation of the Dirac matrices given by

\bea
\bfalpha = \left(\begin{array}{cc} 0 & i\bfsigma \\ -i\bfsigma & 0 \end{array} \right), \qquad \beta = \left(\begin{array}{cc} \v 1_2 & 0 \\ 0 & - \v 1_2 \end{array} \right).
\label{0.2}
\eea
Defining the rising and lowering operators which act on big and small components of spinors in the form

\bea
\Sigma_+ = \left(\begin{array}{cc} 0 & \v 1_2 \\ 0 & 0 \end{array} \right) = \sigma_+ \otimes \v 1_2, \qquad \Sigma_- = (\Sigma_+)^{\dagger}, \qquad \Sigma_3=\beta
\label{0.3}
\eea
and using creation and anhilation operators $\v a = \v r + i \v p, \v a^{\dagger} = \v r - i \v p$ for the oscillator variables, the hamiltonian (\ref{0.1}) can be written in $4 \times 4$ matrix form as

\bea
H = \left(\begin{array}{cc} \v 1_2 m & \bfsigma \cdot \v a^{\dagger} \\ \bfsigma \cdot \v a & - \v 1_2 m \end{array} \right)
\label{0.4}
\eea
or in algebraic form as

\bea
H= \Sigma_+ \v S \cdot \v a + \Sigma_- \v S \cdot \v a^{\dagger} + m \Sigma_3, 
\label{0.5}
\eea
where the spin of the particle is given by $\v S = \v 1_2 \otimes \bfsigma$. In (\ref{0.5}) we have made explicit the separation of the spin and the generators $\Sigma_{\pm}, \Sigma_3$ of $SU(2)$ acting on big and small components of spinors (or particle and anti-particle solutions under the appropriate rotation). We shall refer to $\bfSigma=(\Sigma_+ + \Sigma_-, i(\Sigma_+ - \Sigma_-), \Sigma_3)$ as $*-$spin. It is worth to mention that the projection of the $*-$spin operator $\bfSigma$ is not only related to quasi-relativistic and ultrarelativistic components of states. When combined with oscillator operators, the $*-$spin projection corresponds to states with positive and negative energies. This can be seen from the Foldy-Wouthuysen transformation of this problem \cite{foldy, moreno}.

 The solutions of the stationary Schroedinger equation associated to (\ref{0.5}) were originally obtained in \cite{DO} by writing 

\bea
\left(\begin{array}{cc} \v 1_2 m & \bfsigma \cdot \v a^{\dagger} \\ \bfsigma \cdot \v a & - \v 1_2 m \end{array} \right) \left( \begin{array}{c} \psi_1 \\ \psi_2 \end{array} \right) = E \left( \begin{array}{c} \psi_1 \\ \psi_2 \end{array} \right)
\label{0.6}
\eea
and recognizing that the second row yields $\psi_2 = (E+m)^{-1} (\bfsigma \cdot \v a) \psi_1$. This in turn can be replaced in the first row of (\ref{0.6}) to give

\bea
 \left( m-2E + [(m-E)/(m+E)] (\bfsigma \cdot \v a) (\bfsigma \cdot \v a)^{\dagger} \right)\psi_1 = 0. 
\label{0.61}
\eea
All this can be derived in a simple way by squaring the hamiltonian in (\ref{0.6}), \ie

\bea
\fl \left(\begin{array}{cc} (\bfsigma \cdot \v a^{\dagger}) (\bfsigma \cdot \v a) + m^2 & 0 \\ 0 & (\bfsigma \cdot \v a) (\bfsigma \cdot \v a^{\dagger})+ m^2 \end{array} \right) \left( \begin{array}{c} \psi_1 \\ \psi_2 \end{array} \right) = E^2 \left( \begin{array}{c} \psi_1 \\ \psi_2 \end{array} \right)
\label{0.7}
\eea
which explains the solvability of the problem, resulting in two decoupled equations where finite and infinite degeneracies appear. For a discussion of supersymmetry related to (\ref{0.6}), see \cite{Frank}. We shall prefer, however, the hamiltonian in form (\ref{0.5}), since it is equally useful in deriving the solutions of (\ref{0.6}) and it will prove helpful in further extensions of this system. The dependence of $H$ on ladder operators as in (\ref{0.5}) shows that $I= \v a^{\dagger} \cdot \v a + \frac{1}{2} \Sigma_3$ is an invariant operator which commutes with the squared total angular momentum $J^2$. Therefore, the eigenstates of $I$ with total angular momentum $j$, projection $m_j$ and radial oscillator number $n$ can be used to evaluate the $2 \times 2$ blocks of $H$ and obtain the stationary solutions. We adopt the notation $|n, (l,s)j, m_j\>$ for oscillator states $|nlm\>$ coupled to spin $s$, where $j$ is the total angular momentum with projection $m_j$, $n$ is the radial oscillator number, $l$ is the orbital angular momentum and $m$ its projection on the $z$ axis. We denote $*-$spinors by $|\pm\>$. A pair of states with angular momentum $j$ and such that $I |\quad \>=(2n+j-1)|\quad \>$ is given by

\bea
\fl |\phi_1 \> = |n, (j-1/2,1/2) j, m_j\> |-\>, \quad |\phi_2 \>=|n-1,(j+1/2,1/2) j,m_j\> |+\>.
\label{0.7.1}
\eea
Another pair of states with the same angular momentum $j$ but with $I |\quad
 \>=(2n+j)|\quad \>$ is

\bea
\fl |\phi_3 \>=|n, (j+1/2,1/2) j, m_j\> |-\>, \quad |\phi_4 \>=|n-1,(j-1/2,1/2) j,m_j\> |+\>.
\label{0.7.2}
\eea
The $2\times 2$ blocks of $H$ obtained from these states can be evaluated. We denote their elements by $H(j,2n+j-1)_{ij}=\< \phi_i | H | \phi_j \>$ for $i,j=1,2$ and $H(j,2n+j)_{ij}=\< \phi_i | H | \phi_j \>$ for $i,j=3,4$. From (\ref{0.7.1}) and (\ref{0.7.2}) one obtains the two following matrices

\bea
\fl H(j,2n+j-1)= \left(\begin{array}{cc} -m & \sqrt{2n} \\ \sqrt{2n} & m \end{array} \right), H(j,2n+j)= \left(\begin{array}{cc} -m & \sqrt{2(n+j)} \\ \sqrt{2(n+j)} & m \end{array} \right)
\label{0.7.3}
\eea
leading to the well known energies $E^2=m^2+2(n+j)$ and $E^2=m^2+2n$. Infinite and finite degeneracies come from these two blocks respectively. In connection with these degeneracies, it should be pointed out that a study of the spectrum of the Dirac oscillator \cite{quesne} reveals a symmetry Lie algebra given by a direct sum of $so(4)$ (a compact part with finite-dimensional unitary irreps) and $so(3,1)$ (a non-compact part with infinite-dimensional unitary irreps).

The discussion on the algebraic structure above can be implemented directly in 1 and 2 spatial dimensions as well. Let us introduce the notation $H^{(d)}$ for the hamiltonian in $d$ dimensions. Take $x=r_1, p=p_1, a=a_1$ for the one-dimensional problem. For two dimensions, consider the chiral ladder operators

\bea
A_R = a_1 + i a_2, \qquad A_L = a_1 - ia_2 = (A_R)^{*}
\label{0.8}
\eea
with the properties $[A_R,A_L]=[A_R,(A_L)^{*}]=0,[A_R,A^{\dagger}_R]=[A_L,A^{\dagger}_L]=4$. The hamiltonians for one and two dimensional Dirac oscillators in original form are

\bea
H^{(1)}=\alpha_1 \left( p+i\beta x \right) + m\beta,
\label{0.9}
\eea
with $\alpha_1 = -\sigma_1, \beta=\sigma_3$ and
\bea
H^{(2)}= \sum_{i=1,2}\alpha_i(p_i+i\beta r_i) +  m \beta,
\label{0.10}
\eea
with $\alpha_1 = -\sigma_2, \alpha_2 = - \sigma_1, \beta=\sigma_3$. These hamiltonians can be cast in algebraic form as

\bea
H^{(1)}=\sigma_+ a + \sigma_- a^{\dagger} + m \sigma_3
\label{0.11}
\eea

\bea
H^{(2)}=\sigma_+ A_R + \sigma_- A^{\dagger}_R + m \sigma_3
\label{0.12}
\eea
which exhibit some similarities with (\ref{0.5}). Both of them have a $2\times2$ structure: The spin is absent in one spatial dimension and $\sigma_{\pm}$ corresponds to $*-$spin, while in two dimensions $\sigma_3$ generates the $U(1)$ spin. The solvability can be viewed again as a consequence of the invariants

\bea
I^{(1)}= a^{\dagger} a + \frac{1}{2} \sigma_3
\label{0.12.1}
\eea
in one dimension, and 

\bea
I^{(2)} = A_R A^{\dagger}_R + \frac{1}{2} \sigma_3, \qquad J_3 = A_R A^{\dagger}_R - A_L A^{\dagger}_L +\frac{1}{2} \sigma_3 
\label{0.12.2}
\eea
in two dimensions. We must note, however, that the two dimensional case exhibits some peculiarities which do not appear in the other cases. The conservation of angular momentum $J_3$ comes from the particular combination of $\bfsigma$ and $A_R$ in $H^{(2)}$, together with the absence of $A_L, A^{\dagger}_L$. This absence is also responsible for the infinite degeneracy of all levels. On the other hand, the three dimensional example is manifestly invariant under rotations due to its dependence on $ \v S \cdot \v a$ and $\v S \cdot \v a^{\dagger} $. Its infinite degeneracy comes from the interplay between the infinitely degenerate operator $(\bfsigma \cdot \v a)(\bfsigma \cdot \v a)^{\dagger}$ and $*-$spin, dividing $(H^{(3)})^2$ in two blocks as in (\ref{0.7}). These observations will be used in the following section.

\section{Solvable extensions in one, two and three dimensions}

Now that we have characterized the solvability of these systems as the reduction of $H^{(d)}$ to $2\times2$ blocks, we may readily extend it to $4\times4$ blocks by introducing appropriate potentials. The corresponding Lorentz invariance of such procedure will be discussed later. Consider a hermitean operator of the form $\Phi(\v r,\v p)$ as the potential to be introduced in the total hamiltonian. One has $H^{(d)} = H^{(d)}_0 + \Phi$, with $H^{(d)}_0$ given by the $d-$dimensional Dirac oscillator. On physical grounds, this corresponds to a bound fermion perturbed by a momentum-dependent potential. The scenario in which such type of potentials can arise is discussed with more detail in further sections. For now, let us assume that $\Phi$ may come from either a scalar theory interacting with the bound fermion through Yukawa couplings, or as the scalar component of a vector potential introduced by the minimal coupling prescription. These two possibilities are not equivalent. We may introduce also an internal group for this field, for example the $SU(2)$ associated to isospin in the Yukawa theory or as the gauge group of a non-abelian field. In either case we consider the fundamental representation (doublets). In the three dimensional case, let us impose full rotational invariance of this field in order to preserve integrability, \ie $[\Phi,J_i]=0$. In general, we shall see that a wide class of operators $\Phi(\v r,\v p)$ allows to retain not only integrability, but also solvability of the resulting model. To simplify the discussion, we linearize the field in the variables $\v r, \v p$ (this approximation will be discussed in section 4). We simply have

\bea
\Phi = \Phi_0 + \Phi_1 \v S \cdot \v a + \Phi^{\dagger}_1 \v S \cdot \v a^{\dagger}
\label{01.1}
\eea
where $\Phi_0, \Phi_1$ are constant operators. We denote by $\v T$ the vector of Pauli matrices in isospin space and the corresponding ladder operators by $T_{\pm}=\ahalf (T_1 \pm iT_2)$. Choosing $\Phi_1, \Phi_0$ properly, we may write

\bea
\Phi = \left( T_+ \v S \cdot \v a + T_- \v S \cdot \v a^{\dagger} + \gamma T_3 \right)
\label{01.2}
\eea
with $\gamma$ a real parameter. We may also write the more general
\bea
\Phi = (A+\Sigma_3 B) \left( T_+ \v S \cdot \v a + T_- \v S \cdot \v a^{\dagger} + \gamma T_3 \right)
\label{01.3}
\eea
parametrizing the two possibilities for $\Phi$ mentioned above through the real constants $A,B$. For $A=0$ we have a field which corrects the rest mass of the fermion, while $B=0$ represents the $SU(2)$ potential $A_{\mu}=(\Phi,0,0,0)$. See section 4.

Lower dimensional examples should follow the same pattern to preserve integrability. Thus, we propose the extensions

\bea
\fl H^{(1)}= \sigma_+ a + \sigma_- a^{\dagger} + m \sigma_3 + (A+\sigma_3 B) \left( T_+ a +T_ - a^{\dagger} + \gamma T_3 \right)
\label{01.41}
\eea
\bea
\fl H^{(2)}= \sigma_+ A_R + \sigma_- A^{\dagger}_R + m \sigma_3 + (A+\sigma_3 B) \left( T_+ A_R + T_- A^{\dagger}_R + \gamma T_3 \right)
\label{01.42}
\eea
\bea
\fl H^{(3)}= \Sigma_+ \v S \cdot \v a + \Sigma_- \v S \cdot \v a^{\dagger} + m \Sigma_3 + (A+\Sigma_3 B) \left( T_+ \v S \cdot \v a + T_- \v S \cdot \v a^{\dagger} + \gamma T_3 \right).
\label{01.43}
\eea
With these extensions, it is evident that the new invariants for one, two and three dimensions are

\bea
I^{(1)}= a^{\dagger} a + \frac{1}{2} \sigma_3 + \frac{1}{2} T_3
\label{01.5}
\eea

\bea
I^{(2)} = A_R A^{\dagger}_R + \frac{1}{2} \sigma_3, \quad J_3+\frac{1}{2} T_3 = A_R A^{\dagger}_R - A_L A^{\dagger}_L+ \frac{1}{2} \sigma_3 + \frac{1}{2} T_3
\label{01.6}
\eea

\bea
I^{(3)}= \v a^{\dagger} \cdot \v a + \frac{1}{2} \Sigma_3 + \frac{1}{2} T_3, \quad \v J = \v a^{\dagger} \times \v a + \v S.
\label{01.7}
\eea
Before discussing the solutions of the Schroedinger equations associated to (\ref{01.41},\ref{01.42},\ref{01.43}), our new hamiltonians deserve some comments. It is important to note that the solvability of the resulting Schroedinger equation for these hamiltonians resides completely on the existence of $I^{(d)}$ (which in this case means superintegrability). For instance, in the three dimensional case one may consider any potential of the form $\Phi=F(T_+ \v S \cdot \v a + T_- \v S \cdot \v a^{\dagger} + \gamma T_3)$ where $F$ admits a power expansion. Evidently, $[I^{(3)},\Phi]=0$. A suitable group of states can be used to evaluate the $4 \times 4$ blocks of $H$. We describe this procedure in subsection 3.1, restricting ourselves to the linear case for simplicity.

Despite the fact that the three extensions (\ref{01.41}, \ref{01.42}, \ref{01.43}) exhibit a similar structure, the two dimensional case is, as before, special. In our strive to preserve integrability, we have introduced a field $\Phi$ which depends only on $A_R$ and $A^{\dagger}_R$, leaving $A_L$ and $A^{\dagger}_L$ out of the game. The main consequence of this construction resides in the permanence of the infinite degeneracy of the system (energies do not depend on $A^{\dagger}_L A_L$), as well as a modification of the conserved angular momentum $J_3$ by the addition of isospin (second equality in (\ref{01.6})). The special nature of the original $H^{(2)}$ is responsible for restricting the type of linear extensions and the resulting spectra. In contrast, the three dimensional extension breaks the infinite degeneracy through $T_{\pm}$ and conserves the angular momenta $J^2, J_3$ without modification. We discuss this through the eigenstates and energies of $H^{(3)}$ in the following section.

\subsection{Solutions of the three dimensional case}

Now we turn to the problem of finding the eigenstates of $H^{(3)}$. For this, we simply consider states which satisfy $I |\quad \> = N |\quad \>, J^2 |\quad \> = j(j+1) |\quad \> $. These are four in number. Then we proceed to evaluate the $4 \times 4$ matrix $H(N,j) \equiv \< \quad|H^{(3)}| \quad \>$. Introducing subscripts $\Sigma$ and $T$ for $*-$spinors and isospinors, we write such states as

\bea
|\phi^{N}_1\> = |n, (j+1/2,1/2) j, m_j\> |-\>_{\Sigma} |-\>_{T} \\ \nonumber
|\phi^{N}_2\> = |n, (j-1/2,1/2) j, m_j\> |-\>_{\Sigma}|+\>_{T} \\ \nonumber
|\phi^{N}_3\> = |n-1,(j-1/2,1/2) j,m_j\> |+\>_{\Sigma} |-\>_{T} \\ \nonumber
|\phi^{N}_4\> = |n-1,(j+1/2,1/2) j, m_j\> |+\>_{\Sigma}|+\>_{T}
\label{02.1}
\eea
where, as before, $n$ is the oscillator radial number, $j$ is the total angular momentum and $m_j$ its projection in the $z$ axis. These are eigenstates of $I^{(3)}$ with eigenvalue $N=2n+j-1/2$. The resulting $4\times 4$ blocks of $H$ with elements $H(N,j)_{kl}=\<\phi^{N}_k|H|\phi^{N}_l\>$ are

\bea
\fl \left( \begin{array}{cccc} -m-(A-B)\gamma & (A-B)\sqrt{2(n+j)} & -\sqrt{2(n+j)} & 0 \\ (A-B)\sqrt{2(n+j)} & -m+(A-B)\gamma & 0 & \sqrt{2n} \\ -\sqrt{2(n+j)} & 0 & m-(A+B)\gamma & (A+B)\sqrt{2n} \\ 0 & \sqrt{2n} & (A+B)\sqrt{2n} & m+(A+B)\gamma \end{array} \right)
\label{02.2}
\eea
and the secular equation $|H(N)-E|=0$ can be solved explicitly using the formula for the roots of a quartic polynomial. However, we shall contempt ourselves with the explicit solutions of the $1+1$ dimensional example in further sections, as the blocks of $H^{(1)}$ and $H^{(3)}$ are quite similar when $j=1/2$. It is also evident that the infinite degeneracy is now broken, since one cannot reduce $H(N)$ to smaller blocks where only $n$ appears. The exception to this ocurrs when $A=B=0$, which obviously recovers the usual Dirac oscillator.

\section{Lorentz covariant formulation}

Here we discuss the specific form of Lorentz covariant external fields producing the solvable extension we have established. Before proceeding, we should mention that the potential $\Phi$ in one and two dimensions indicated in (\ref{01.41},\ref{01.42}), is spin independent. This makes our task easier, since $\Phi$ can be directly related to the external field as shall be indicated. The three dimensional case, however, should be treated separately due to the presence of $\v S$ in the potential $\Phi$ considered in (\ref{01.43}). Therefore, we shall consider three alternatives for generating $\Phi$: a parity violating coupling to an external non-local vector field (mediated by $\gamma_5$), a Pauli coupling to a non-local field tensor and a Yukawa coupling to a non-local scalar field. All of them shall have an internal symmetry group $SU(2)$ as indicated before. The alternatives involving a non-local gauge field can be brought together when linear approximations of fields are considered, as we shall see.

\subsection{3+1 dimensional case}

Here we use the metric tensor $\eta = {\rm diag\ }(1,-1,-1,-1)$. The original covariant formulation of the Dirac oscillator \cite{marcosbook} considered a four vector $u$ such that, in some inertial frame, $u_{\mu}=(1,0,0,0)$. For the Dirac oscillator of many particles, such vector was related to the center of mass of the system. In the present case, the meaning of $u$ is given through the introduction of the anomalous (Pauli) coupling to an external field tensor. The equation leading to the Dirac oscillator hamiltonian (\ref{0.1}) is

\bea
[\gamma_{\mu}(p^{\mu}+i\gamma_{\nu}u^{\nu}r^{\mu}_{\perp}) + m ] \psi = 0,
\label{uno}
\eea
where $r^{\mu}_{\perp}=r^{\mu}-(r^{\nu}u_{\nu})u_{\mu}$ and 

\bea
\gamma_j = \left( \begin{array}{cc} 0 & i \sigma_j \\  i \sigma_j & 0 \end{array} \right), \quad \gamma_0 = \left( \begin{array}{cc} \v 1_2 & 0 \\ 0 & -\v 1_2 \end{array} \right).
\label{uno.1}
\eea
Eq.(\ref{uno}) can be cast as 

\bea
[\gamma_{\mu}p^{\mu} + m + S_{\mu \nu}F^{\mu \nu}] \psi = 0
\label{uno.2}
\eea
with $F^{\mu \nu} = u^{\mu}r^{\nu} - u^{\nu}r^{\mu}$ and $S_{\mu \nu}= i [\gamma_{\mu},\gamma_{\nu}]$. Thus, the meaning of $u$ is obtained by means of the divergence of the Faraday tensor as $\partial_{\mu} F^{\mu \nu} = -u^{\nu}$, \ie $u$ is an external current. It must be mentioned that another possibility to write a covariant form of the interaction comes through the dual tensor. The anomalous coupling with an electric field in the form $\gamma_5 S_{\mu \nu} \tilde{F}^{\mu \nu}$ with $\tilde{F}^{\mu \nu}=\epsilon^{\mu \nu \alpha \beta } F_{\alpha \beta}$, produces the same result.

With the aid of the vector $u_{\mu}$ we can introduce more interactions in a covariant way. As announced before, a non-local, non-abelian field tensor $\fcal^{\mu \nu}=\sum_{i=1}^{3} T_i \fcal^{\mu \nu}_i$ can be introduced in the equation by means of the Pauli coupling. The theory underlying this type of field can be found in many references \cite{nonlocal1}, \cite{nonlocal2} and was motivated by the possibility of introducing high order derivatives in field actions. From all these possibilities, we choose the path followed in \cite{nonlocal3} by considering that the field acts on wave functions as a bilocal kernel, \ie it can be represented as an operator depending on $r_{\mu}, p_{\mu}$ (but not $\v S$). The integro-differential equations obeyed by $\fcal^{\mu \nu}$ are also in \cite{nonlocal3} and can be used to elucidate the nature of the external source generating such field. We shall proceed to find the external current $j_{\mu}$ producing such fields after we give the expressions for the interactions. Let us now consider a linearization of the field in the $p,r$ variables: an approximation which corresponds to the first corrections due to internal structure of particles described by the non-abelian field. This means that both wavelength and particle structure are small compared to the scale of the system, which is dictated by the oscillator frequency. The expansion of the field in $r_{\mu},p_{\nu}$ can be truncated by keeping linear terms only. We propose

\bea
\fcal_1^{\mu \nu} = \epsilon^{\mu \nu \lambda \rho} u_{\lambda} r_{\perp \rho} \\
\fcal_2^{\mu \nu} = \epsilon^{\mu \nu \lambda \rho} u_{\lambda} p_{\perp \rho} \\
\fcal_3^{\mu \nu} = 0,
\label{dos}
\eea
for which the Dirac equation reads
\bea
[\gamma_{\mu}p^{\mu} + m + S_{\mu \nu}F^{\mu \nu} + B S_{\mu \nu}\fcal^{\mu \nu}] \psi = 0,
\label{tres}
\eea
This equation reduces to the hamiltonian form given in (\ref{01.43}) with $A=\gamma=0$, as can be verified by setting $u_{\mu}=(1,0,0,0)$. To produce a term $\gamma \Sigma_3 T_3$ in the hamiltonian, we simply recognize that such a potential corrects the rest mass of the Dirac particle. A non-abelian scalar field can be introduced for this purpose in analogy with the one and two dimensional cases explained below. Notably, one could also consider the dual tensor with components $\tilde{\fcal}_i^{\mu \nu}=\epsilon^{\mu \nu}_{\,\, \beta \alpha} \tilde{\fcal}_i^{\beta \alpha}$ and an interaction of the form 
$\gamma_5 S_{\mu \nu} \tilde{\fcal}^{\mu \nu}$ in order to produce
(\ref{01.43}).

The other case, \ie $B=0, A =1$, comes from considering again a parity violating coupling when we use the minimal coupling prescription. Such a form is encountered in field theories where chirality plays an important role \cite{weinberg}, e.g. the electroweak model. Again, we need non-locality in our fields. By using a non-abelian 4-potential (e.g. a current vector with isospin 1/2) of the form $A^{\mu}= \sum_{i=1}^{3} A^{\mu}_i T_i$, where

\bea
A^{\mu}_1 = r_{\perp \mu}, \quad A^{\mu}_2 = p_{\perp \mu}, \quad A^{\mu}_3 = 0,
\label{cuatro}
\eea
we can see that the equation

\bea
[\gamma_{\mu}p^{\mu} + m + S_{\mu \nu}F^{\mu \nu} + A \gamma_5 \gamma_{\mu}A^{\mu}] \psi = 0
\label{cinco}
\eea
produces the desired potential in the hamiltonian for $B=\gamma=0$. The term with $T_3$ does not appear in this case, but can be generated by the unitary transformation $\psi' = \exp(i T_3 u_{\nu} r^{\nu} ) \psi$. This leaves $p_{\perp}$ invariant, while the matrices $T_1, T_2$ in the hamiltonian simply rotate.

The nature of such field can be elucidated by inserting our $\fcal_{\mu \nu}$ in the corresponding non-local field equations and finding the external current producing it. For instance,
using $\tilde{\fcal}^{\mu \nu}$ one has $\tilde{\fcal}^{\mu \nu}= u^{\mu}(r^{\nu}_{\perp}T_1+p^{\nu}_{\perp}T_2) - \mu \leftrightarrow \nu$ and its expression in terms of a 4-potential (see (12) in \cite{nonlocal3}) reads

\bea
\tilde{\fcal}^{\mu \nu} = i([p^{\mu},B^{\nu}]-\mu \leftrightarrow \nu) + [B_{\mu},B_{\nu}].
\label{cinco.1}
\eea
This yields $B_{\mu}$ up to gauge transformations as

\bea
B_{\mu}=u_{\mu} (\frac{1}{2}  r_{\mu} r^{\nu}_{\perp} T_1 + r_{\nu} p^{\nu}_{\perp} T_2 ) \qquad \mbox{Bilinear in $p$, $r$}.
\label{cinco.2}
\eea
The field equations give the current as

\bea
j^{\nu} = i[p_{\mu},\tilde{\fcal}^{\mu \nu}] + [B_{\mu},\tilde{\fcal}^{\mu \nu}] \\ \nonumber
=-u^{\nu} T_1 + p^{\nu}_{\perp} + \left( \frac{1}{2} \lbrace p_{\perp}^{\nu}, r_{\mu} r^{\mu}_{\perp} \rbrace - \lbrace p_{\perp}^{\mu}, r_{\perp}^{\nu} \rbrace r_{\mu} \right) T_2 \\ \nonumber
=-u^{\nu} T_1 + p^{\nu}_{\perp} + \mbox{ trilinear terms in $p$,$r$ }.
\label{cinco.3}
\eea
So far, no approximations have been made. If we focus on the independent and linear terms we find a) a constant current $u_{\mu}$ as in the Dirac oscillator case, b) a spatial part given by the momentum of the particle, as if the imposed external current would 'follow' the electron.

In order to put the cases $A \neq 0, B \neq 0$ together, we may consider a 4-potential given by the combination $A A_{\mu} + B B_{\mu}$, with $A_{\mu}$, $B_{\mu}$ given above. The linearity in $p$ and $r$ invoked in our argumentation (e.g. in the case where only the first non-local corrections are considered and field intensities are small compared to the oscillator frequency) implies that the interaction $\gamma_5 \gamma^{\mu}(A A_{\mu}+B B_{\mu})$ can be approximated by $A \gamma_5 \gamma^{\mu}A_{\mu}$ when inserted in the Dirac oscillator equation. The anomalous coupling $\gamma_5 S_{\mu \nu} \tilde{\fcal}^{\mu \nu}$ with the combination of $A_{\mu}$ and $B_{\mu}$ yields the one produced by $B_{\mu}$ when we ignore quadratic terms in $p,r$ and in the constants $A,B$. We would like to stress that despite the fact that the linear approximation is helpful to put both cases together, one may consider each case separately without approximations in the light of
(\ref{cinco.1},\ref{cinco.2}).

The physical origin of our potentials, as we have seen, depends strongly on the validity of non-local theories and the possible existence of extended particles as discussed by Yukawa (composite particles described effectively by such fields is another possibility).

\subsection{2+1 dimensional case}

Here we construct a field producing the two dimensional hamiltonian (\ref{01.42}). Greek indices now run from 0 to 2. Since spin does not appear in the resulting potential $\Phi$, we identify it with a non-local scalar field, according to Yukawa's formulation in \cite{yukawa1}, \cite{yukawa2}. In the presence of isospin $\v T$, the potential has an expansion of the form

\bea
\Phi = \bfphi \cdot \v T = \sum_{j=1}^{3} T_j \int d^4k^{(j)} d^4l^{(j)} f_j(k,l) e^{ik^{(j)}_{\mu} x^{\mu}} e^{il^{(j)}_{\mu} p^{\mu}},
\label{A.2}
\eea
where $k^{(i)}$ are 2+1 dimensional Lorentz vectors representing the momentum of a scalar particle of mass $M$, while the $l^{(i)}$ vectors represent the transversal {\it radius.\ } We shall consider only one Fourier component. According to Yukawa \cite{yukawa2}, the corresponding 4-vectors obey

\bea
k^{(1)}_{\mu} k^{(1) \mu} = k^{(2)}_{\mu} k^{(2) \mu} = M^2 \\
l^{(1)}_{\mu} l^{(1) \mu} = l^{(2)}_{\mu} l^{(2) \mu} = -\lambda^2 \\
k^{(1)}_{\mu} l^{(1) \mu} = k^{(2)}_{\mu} l^{(2) \mu} = 0.
\label{seis}
\eea
where $\lambda$ is the radius of the scalar particle. Again, linearization in $p$ will be necessary and it is related to the first corrections to a scalar particle (or a doublet of particles, since we use $SU(2)$ as isospin) with small $\lambda$. Now, we can choose a frame of reference where the set of vectors above is given by

\bea
k^{(1)}_{\mu} = (\sqrt{(k^{(1)})^2+M^2},k^{(1)},0) \qquad l^{(1)}_{\mu} = (0,0,\lambda) \\
k^{(2)}_{\mu} = (\sqrt{(k^{(2)})^2+M^2},0,k^{(2)}) \qquad l^{(2)}_{\mu} = (0,\lambda,0).
\label{siete}
\eea
and define 

\bea
\kappa^{(1)}_{\mu} = \left(\frac{k^{(1)}_{\nu} l^{(2) \nu}}{\lambda} \right) l^{(2)}_{\mu} \qquad \kappa^{(2)}_{\mu} = \left(\frac{k^{(2)}_{\nu} l^{(1) \nu}}{\lambda} \right) l^{(1)}_{\mu}.
\label{ocho}
\eea
Finally, we give the field $\Phi = \bfphi \cdot \v T$ as

\bea
\phi_1 = \kappa^{(1)}_{\mu} r^{\mu} - l^{(1)}_{\mu} p^{\mu}, \quad \phi_2 = \kappa^{(2)}_{\mu} r^{\mu} - l^{(2)}_{\mu} p^{\mu}, \quad \phi_3 = \gamma.
\label{nueve}
\eea
The equation

\bea
[\gamma_{\mu}(p^{\mu}+i\gamma_{\nu}u^{\nu} r^{\mu}_{\perp}) + m + \Phi] \psi = 0,
\label{diez}
\eea
with $\gamma_0= \sigma_3, \gamma_1=i \sigma_1, \gamma_2= -i \sigma_2$, yields the two dimensional hamiltonian in (\ref{01.42}) for $A=0,B=1$. The frame of reference where this is possible, is dictated by the unitary time-like vector $u$ chosen as $u_{\mu}= \widehat{k^{(2)}_{\mu} - \kappa^{(2)}_{\mu}} = \widehat{k^{(1)}_{\mu} - \kappa^{(1)}_{\mu}}$. The case $A=1,B=0$ is obtained by considering a non-abelian gauge field whose scalar component is written in the same spirit, \ie 

\bea
[\gamma_{\mu}(p^{\mu}+i\gamma_{\nu}u^{\nu} r^{\mu}_{\perp}+A_{\mu}) + m ] \psi = 0 \\
A_{\mu}= u_{\mu} \Phi
\label{once}
\eea

\subsection{1+1 dimensional case}

We proceed along the same line by introducing a non-local scalar field. Greek indices now run from 0 to 1 and Dirac matrices are $\gamma_0=\sigma_3, \gamma_1 = -i \sigma_2$. We use the momentum and radius vectors $k_{\mu},l_{\mu}$ in a frame of reference such that

\bea
k_{\mu}=(m,0), \qquad l_{\mu}=(0,\lambda),
\label{doce}
\eea
define $\kappa_{\mu} = (\frac{M}{\lambda}) l_{\mu}$ and set the field components as

\bea
\phi_1 = \kappa_{\mu} r^{\mu}, \qquad \phi_2 = l_{\mu} p^{\mu}, \qquad \phi_3 = \gamma.
\label{trece}
\eea
Using the same Yukawa coupling and choosing $u_{\mu}= \hat k_{\mu}$ we obtain the desired potential in the hamiltonian for $A=0$. The case $B=0$ is obtained as before, \ie by using a gauge potential $ A_{\mu}= u_{\mu} \Phi $.

\section{Detailed analysis of $1+1$ dimensional extensions}

Now we concentrate on the $1+1$ dimensional case for simplicity. Furthermore, we shall find a similarity between the block (\ref{02.2}) for $j=1/2$ and the corresponding block for the one dimensional problem. Let us write the Dirac equation in hamiltonian form by choosing $u_{\mu}=(1,0)$. It is useful to introduce a coupling constant $\alpha$ for the field components $\phi_1, \phi_2$ in (\ref{trece}). The resulting Schroedinger equation in terms of ladder operators reads

\bea
\fl \left[ \sigma_+ a + \sigma_- a^{\dagger} + m \sigma_3 + \left(A \sigma_3 + B \right) \left( \alpha T_+ a + \alpha T_- a^{\dagger} + \gamma \right) \right] \psi = i \frac{\partial \psi}{\partial t},
\label{6}
\eea
for which the usual Yukawa coupling is obtained when $A=1$ and $B=0$, while for $A=0$, $B=1$ we obtain the zero component of the non-local potential. Now we proceed to find the solutions of the stationary version of (\ref{6}) by noting that the operator

\bea
I = a^{\dagger}a +\frac{1}{2}(\sigma_3 + T_3) - 1
\label{7}
\eea
satisfies $[H,I]=0$, \ie $I$ is a conserved quantity and it gives the set of states for which $H$ is reduced to a block diagonal form. Such states are formed by harmonic oscillator states $|n\>$ and states $|\tau \sigma \>$ for isospin projection $\tau = \pm $ and $*-$spin projection $\sigma = \pm $. The states are given by
\bea
|\phi_1^n\> = |n+2\> |--\>
\quad\quad
|\phi_2^n\> = |n+1\> |-+\> \nonumber\\
|\phi_3^n\> = |n+1\> |+-\>
\quad\quad
|\phi_4^n\> = |n\> |++\>
\label{8}
\eea
with $n \geq 0$ and satisfy

\bea
I |\phi_i^n \> = n |\phi_i^n \>, \qquad i=1,...,4.
\label{9}
\eea
The Hamiltonian $H$ in (\ref{6}) has a matrix representation
\bea
H=\left(
\begin{array}{cccc}
H_0&0&0&\dots\\
0&H_1&0&\dots\\
0&0&H_2&\\
\vdots&\vdots&&\ddots
\end{array}
\right),
\label{10}
\eea
where the diagonal blocks $H_{n}$ have matrix elements $(H_n)_{ij}=\< \phi_i^n | H | \phi_j^n \>$. They have the form

\bea
\fl H_{n}=\left(
\begin{array}{cccc}
-m-(B-A)\gamma&\alpha(B-A)\sqrt{n+2}&\sqrt{n+2}&0\\
\alpha(B-A)\sqrt{n+2}&-m+(B-A)\gamma&0&\sqrt{n+1}\\
\sqrt{n+2}&0&m-(B+A)\gamma&\alpha(B+A)\sqrt{n+1}\\
0&\sqrt{n+1}&\alpha(B+A)\sqrt{n+1}&m+(A+B)\gamma
\end{array} 
\right), \label{11} \nonumber \\
\eea
except for the lowest states forming a singlet and a triplet which we now indicate. 

The state $|0,--\>$ forms a singlet of $I$, \ie the only state with eigenvalue equation $I|0,--\>=-2|0,--\>$. Its energy is given by $H|0,--\>=-(m+(B-A)\gamma)|0,--\>$ trivially. The states $|1,--\>,|0,+-\>,|0,-+\> $ form a triplet of $I$ with eigenvalue $-1$, \ie $I |1,--\> = -|1,--\>$ (similarly for the other two). They give a block of the form

\bea
H_{-1}\equiv\left(
\begin{array}{ccc}
-m-(B-A)\gamma&\alpha(B-A)&1\\
\alpha(B-A)&-m+(B-A)\gamma&0\\
1&0&m-(B+A)\gamma
\end{array}
\right)
\label{11.1}
\eea
leading to the lowest energy levels as solutions of a cubic polynomial. These two cases represent corrections to the vacuum energies, as can be seen by choosing vanishing interaction parameters $A,B$. In the following, we shall concentrate on general states described by (\ref{8}), \ie states satisfying (\ref{9}) for values $n \geq 0$. The spectrum can be obtained from the $4$ dimensional secular equation $|H_n-E\v 1_4|=0$ using (\ref{11}), with $\v 1_4$ the $4 \times 4$ identity matrix. This, of course, lies on the fact that the formulae for the roots of a fourth degree polynomial are known \cite{quartic}. Eigenvectors can be written explicitly as well.

For the sake of simplicity we study particular cases and discuss their physical importance.

\subsection{The case $A=0$, $B=1$}

This case corresponds to gauge fields acting on the Dirac particle. Conversely, one could put $B=0, A=1$, with analogous solutions except for the presence of some signs in the parameters $\alpha, \gamma$. The situation is then quite general from the point of view of the solutions. The resulting eigenvalues (\ref{ener}) exhibit a dependence on $n$ such that the factors in front of it get corrected by the presence of $\alpha, \gamma$. The oscillator vacuum energy $n=0$ is modified as well. Therefore, both frequency and rest mass acquire corrections due to the interaction, as expected. The blocks of $H$ are given by 

\bea
H_{n}=\left(
\begin{array}{cccc}
-m-\gamma&\alpha\sqrt{n+2}&\sqrt{n+2}&0\\
\alpha\sqrt{n+2}&-m+\gamma&0&\sqrt{n+1}\\
\sqrt{n+2}&0&m-\gamma&\alpha\sqrt{n+1}\\
0&\sqrt{n+1}&\alpha\sqrt{n+1}&m+\gamma
\end{array}
\right).
\label{block}
\eea
The following definitions are helpful
\bea
p=&\frac{2}{3} \left((2 n+1) 
\left(1+\alpha ^2\right)+2 \gamma ^2+2m^2\right)\nonumber\\
q=&16 m^2\left(m^2-\left(1+\gamma ^2\right)\right)+
\left(1+4 \gamma ^2\right)^2+2\left(1+4 m^2-8 \gamma ^2\right) \alpha ^2+
\alpha ^4-\nonumber\\
&16\left(1+m^2 \left(\alpha ^2-2\right)+\gamma ^2-\alpha ^2 
\left(1+2 \gamma ^2\right)+\alpha ^4\right)(n+2)+\nonumber\\
&16\left(1- \alpha ^2+ \alpha ^4\right) (n+2)^2\nonumber\\
r=&108\Big( 
\left((n+2)\left(1- \alpha ^2\right)+m^2-\gamma ^2\right) 
\left((n+1)\left(1-\alpha ^2\right)+m^2-\gamma ^2\right)p\nonumber\\ 
&-\frac{1}{16} p^3
+\left(m \alpha ^2+\gamma \right)^2\Big)
\nonumber\\
s=&
\frac{q}{3}
\left(\frac{2}{ r+\sqrt{r^2-4 q^3}}\right)^{1/3}+
\frac{1}{3}
\left(\frac{r+\sqrt{r^2-4
 q^3}}{2}\right)^{1/3}.
\eea
The eigenvalues can be expressed as:
\bea
E_i(n)=\left\{
\begin{array}{l}
\frac{\sqrt{p+s}}{2}+
\frac{(-1)^i}{2}\sqrt{2 p-s-4\frac{m\alpha^2+\gamma}{\sqrt{p+s}}},
\qquad i=1,2 \\ \\
-\frac{\sqrt{p+s}}{2}+
\frac{(-1)^i}{2}\sqrt{2 p-s+4\frac{m\alpha^2+\gamma}{\sqrt{p+s}}},
\qquad i=3,4
\end{array}\right.
\label{ener}
\eea
We denote the eigenvectors of (\ref{block}) without normalization by $V_i$ such that $H_n V_i= E_{i}(n)V_i$. Their components $V_{ji}$ for $j=1,...,4$ can be written explicitly as
\bea
V_{1i}=&  \alpha\sqrt{n+2}\left( 
(n+1)(1-\alpha^2)+(m-E_i)^2-\gamma^2\right)\nonumber\\
V_{2i}=
&(m+\gamma ) \left((n+2)(1-\alpha ^2)+\alpha^2+m^2-\gamma ^2\right)-\nonumber\\
&\left((n+2)(1+\alpha ^2)-\alpha^2+(m+\gamma)^2\right) 
E_i+
(\gamma-m) E_i^2+E_i^3\nonumber\\
V_{3i}=&\alpha\left((2n+1)E_i-m-\gamma \right)\nonumber\\
V_{4i}=&
\sqrt{n+1} \left((n+2)(\alpha ^2-1)+(\gamma +E_i)^2-m^2\right).
\eea

\subsection{The case $A=0, B\alpha=1, m=\gamma$}

We put this case separately, although it can be reached from the solutions in the last subsection.
For this set of parameters, it is evident that a new symmetry arises in the form of permutational invariance $\bfsigma \leftrightarrow \v T$ and conservation of the total spin magnitude $(\bfsigma+\v T)^2$. In the context of particle physics, such a conservation law indicates that isospin $+$ fermion charge is a good quantum number. The characteristic polynomial is factored in qubic and linear expressions as a result of the antisymmetric singlet and the symmetric triplet of the total spin $\bfsigma + \v T$. In fact, the cubic part is a {\it depressed\ }cubic. The spin zero state is infinitely degenerate with zero energy. We use the states

\bea
\fl |\chi_1^n\> \equiv \frac{1}{\sqrt{2}}|n+1\> (|-+\>-|+-\>)= |n+1\>|0,0\>,
\quad
|\chi_2^n\> \equiv |n+2\> |--\>=|n+2\>|1,-1\>, \nonumber\\
\fl |\chi_3^n\> \equiv \frac{1}{\sqrt{2}}|n+1\> (|-+\>+|+-\>)=|n+1\>|1,0\>,
\quad
|\chi_4^n\> \equiv |n\> |++\> = |n\>|1,1\>
\label{16}
\eea
where $|s,m_s\>$ are states of the total spin $\bfsigma + \v T$ with $s=0,1$ and its projection $m_s=-1,0,1$. It is straightforward to compute $(H_n)_{ij}=\< \chi_i^n | H | \chi_j^n \>$. We obtain

\bea
 H_{n}=\left(
\begin{array}{cccc}
0&0&0&0\\
0&2\gamma&\sqrt{2(n+1)}&0\\
0&\sqrt{2(n+1)}&0&\sqrt{2(n+2)}\\
0&0&\sqrt{2(n+2)}&-2\gamma
\end{array}
\right)
\label{17}
\eea
This yields eigenvalues which are similar to those for the previous case, but with $\alpha=1, m=\gamma$.

\subsection{The case $\alpha = 0$}

As a simple application of our relativistic model, one can eliminate the presence of the field, except for its static part. For a perturbed confined particle, this case only takes into account a static correction to its rest mass and zero point energy. As we shall see, the oscillator number is not affected by factors involving $\gamma$, leaving the Dirac oscillator frequency untouched. One could state that the simplest case of mass generation (even at the level of classical fields) is thus achieved. The matrix becomes

\bea
 \fl H_{n}=\left(
\begin{array}{cccc}
-m-(B-A)\gamma&0&\sqrt{n+2}&0\\
0&-m+(B-A)\gamma&0&\sqrt{n+1}\\
\sqrt{n+2}&0&m-(B+A)\gamma&0\\
0&\sqrt{n+1}&0&m+(A+B)\gamma
\end{array}
\right),
\label{17.2}
\eea
with eigenvalues

\bea
\fl E_1(n) = -B \gamma - \sqrt{n+2+(m-A\gamma)^2}, \qquad E_2(n) = -B \gamma + \sqrt{n+2+(m-A\gamma)^2} \nonumber \\
\fl E_3(n) = B \gamma - \sqrt{n+1+(m+A\gamma)^2}, \qquad E_4(n) = B \gamma + \sqrt{n+1+(m+A\gamma)^2}.
\label{17.3}
\eea
The eigenvectors of $H_n$ are given by the rows of the following matrix
\bea
 V^T=\left(
\begin{array}{cccc}
\frac{E_1(n)+(A+B)\gamma-m}{\sqrt{n+2}}\gamma&0&1&0\\
\frac{E_2(n)+(A+B)\gamma-m}{\sqrt{n+2}}\gamma&0&1&0\\
0&\frac{E_3(n)-(A+B)\gamma-m}{\sqrt{n+1}}\gamma&0&1\\
0&\frac{E_4(n)-(A+B)\gamma-m}{\sqrt{n+1}}\gamma&0&1
\end{array}
\right).
\label{17.4}
\eea

\subsection{The case $m=\gamma=0$}

This corresponds to a confined massless fermion perturbed by a field with no static part. Although one could eliminate direct corrections to the rest mass by setting $B=0$ in the hamiltonian, the energies for this case contain terms analogous to the rest energies due to $\alpha$ and with a two-fold splitting. The hamitonian is now independent of $T_3,\sigma_3$ and is invariant under the map
\bea
\bfsigma \mapsto -\bfsigma, \quad \v T \mapsto -\v T, \quad x \mapsto -x, \quad p \mapsto -p.
\label{17.5}
\eea
The characteristic polynomial of each block $H_n$ is now biquadratic. The four eigenvalues are found by choosing one of the alternating signs in the following expression

\bea
E(n)= \pm \sqrt{\frac{b(n) \pm a(n)}{2}}
\label{17.6}
\eea
where

\bea
\fl a(n) = \left\{ \left[16B^2\alpha^2(A^2\alpha^2+1)\right] (n+2)^2 - \left[8B\alpha^2(A(A+B)^2\alpha^2+A+2B) \right](n+2) \right. \\ + \left. \left[1+(A+B)^2\alpha^2 \right]^2 \right\}^{1/2} \nonumber \\
\fl b(n) = 2\left[1+(A^2+B^2)\alpha^2\right]n + 3 \left[1+(A+B)^2\alpha^2 \right].
\label{17.7}
\eea
Eigenvectors can be obtained easily.

\section{Entanglement of a confined particle with the external field}

\begin{center}
  \begin{figure}[ht!]
    \includegraphics[scale=.65]{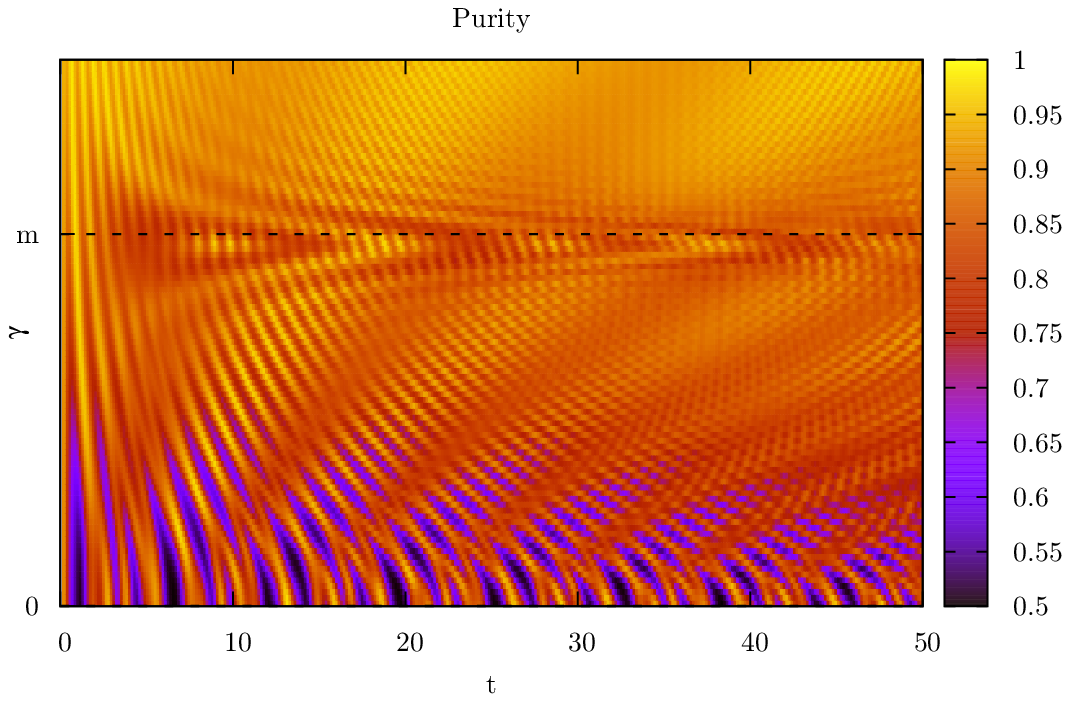}
    \includegraphics[scale=.65]{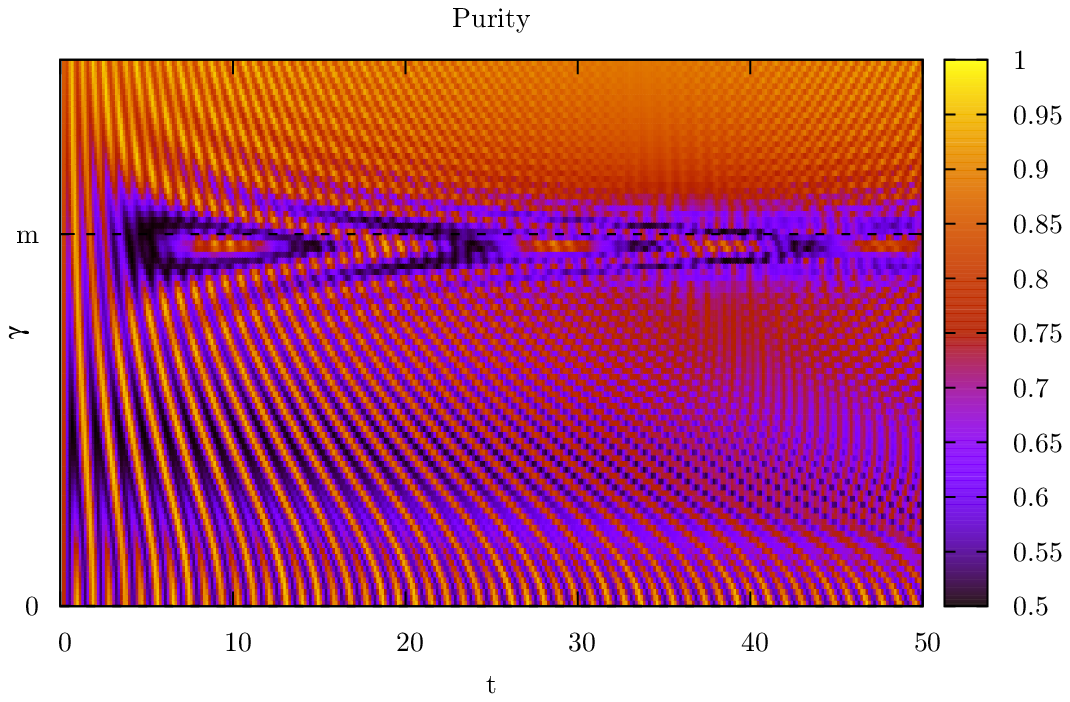}
    \includegraphics[scale=.65]{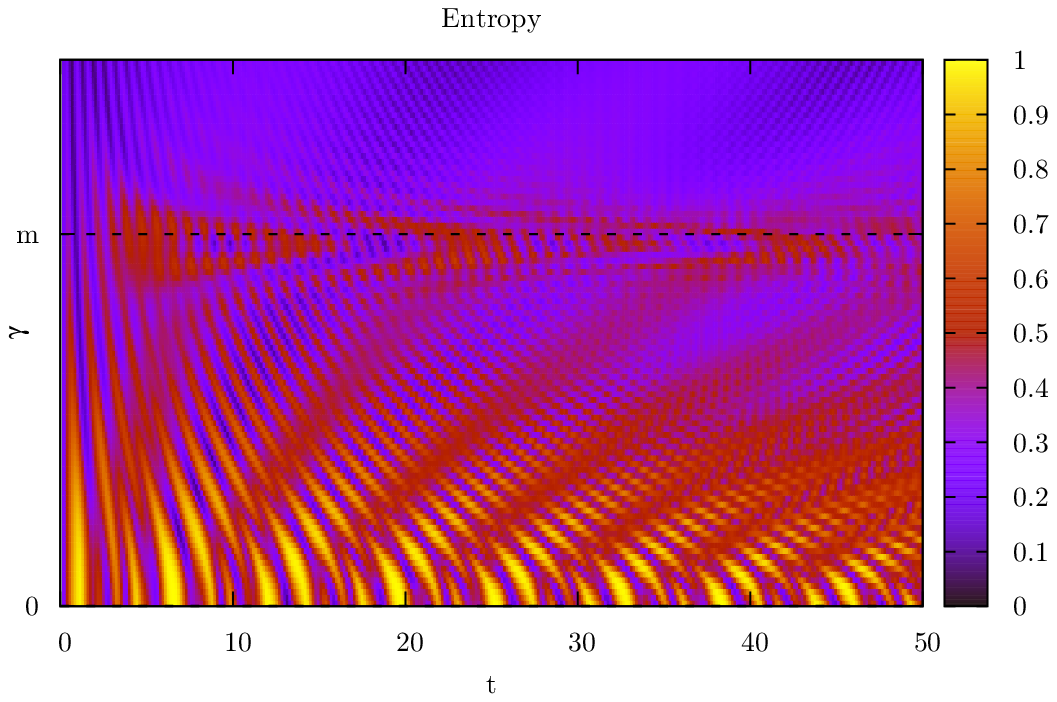}
    \includegraphics[scale=.65]{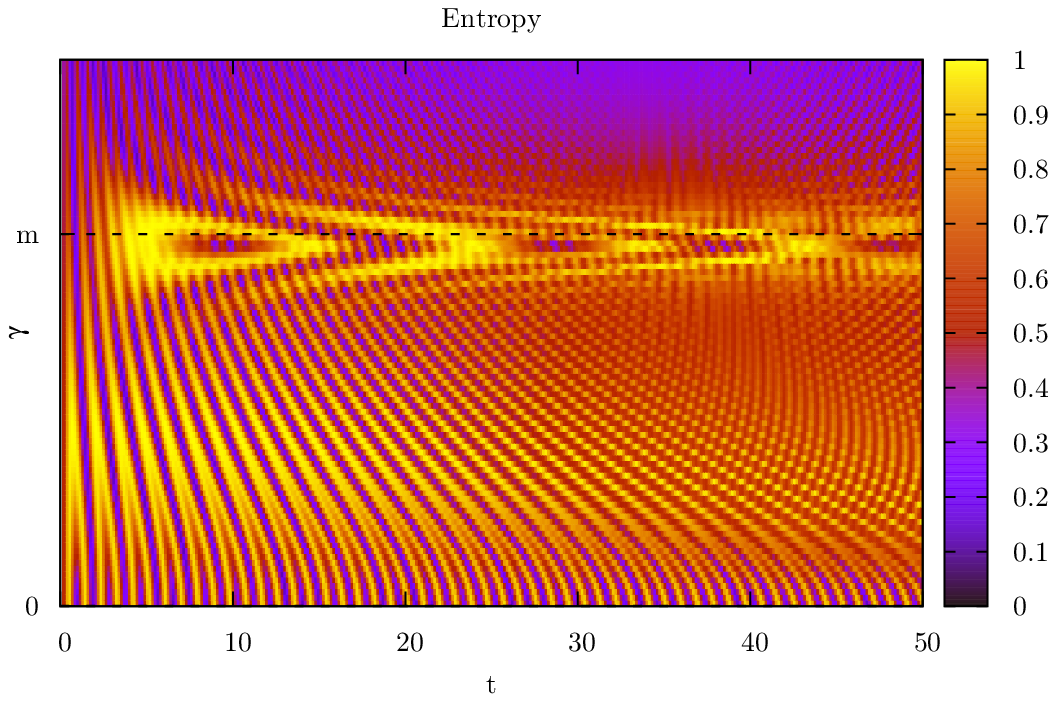}
\caption{Purity and Entropy of the Dirac oscillator state as functions of time $t$ (abscissa)
and static field $\gamma$ (ordinate). Parameters are $m=3.2$, $\alpha=1.2$. The left column corresponds to $\theta=\pi/4$ and the right column to $\theta=0$. We take $n=0$ for the Dirac Oscillator state. The structure at $\gamma \sim m$ is present in all density plots and is persistent at all times. In this region we observe an entropy increase and a purity decrease. This is an indication of maximal entanglement between the external field and the positive and negative energy states of the confined fermion. The situation corresponds to a regime in which the field intensity is comparable to the rest energy: particles can be created in the context of quantum field theory.}
    \label{fig2}
  \end{figure}
\end{center}

As an example for dynamics in this system we shall now study the evolution of entanglement of the Dirac oscillator and the field. We assume our Dirac particle to be initially in 
some eigenstate of the Dirac oscillator, e.g. of positive energy \cite{marcosbook} given by

\bea
|\chi_n\> = 
A_n^{(+)} |n\>|+\> + A_n^{(-)} |n+1\>|-\>,
\label{4.1}
\eea
where
\bea
A_n^{(+)}&=&\sqrt{\frac{n}{n+(m-\epsilon_n)^2}}\nonumber\\
A_n^{(-)}&=&\frac{m-\epsilon_n}{\sqrt{n+(m-\epsilon_n)^2}}\nonumber\\
\epsilon_n&=&\sqrt{2n+1+m^2}
\eea
The complete system is considered to be initially in a factorized state of the form Dirac oscillator $\times$ Isospin, \ie $\psi=\chi_n 
\otimes \chi$, $\chi$ being a Pauli spinor for isospin. We put in general 
$|\chi\>=1/\sqrt{2}(\cos{\theta}|+\>+\sin{\theta}|-\>)$
and examine two different angles $\theta=0,\pi/4$. We shall see that the 
particular choice of $\theta $ will not be relevant to what we want to 
show. The initial condition can be written as linear combinations of the states (\ref{8}) with coefficients $K_i$, $i=1,...,4$. Such states, in turn, are expressed as superpositions of eigenstates of the system $|\psi_n^{j}\>$. This results in

\bea
|\psi \>&=& K_1 |\phi_{n+1}^{1}\> + K_3|\phi_{n+1}^{3}\> \nonumber \\& & K_2 |\phi_{n+2}^{2}\> + K_4 |\phi_{n+2}^{4}\> \nonumber \\
&=& K_1 \sum_{j=1}^{4} D_{1,j}^{(n+1)} |\psi_{n+1}^{j}\> + 
    K_3 \sum_{j=1}^{4} D_{3,j}^{(n+1)} |\psi_{n+1}^{j}\> \nonumber\\
&+& K_2 \sum_{j=1}^{4} D_{2,j}^{(n+2)} |\psi_{n+2}^{j}\> + 
    K_4 \sum_{j=1}^{4} D_{4,j}^{(n+2)} |\psi_{n+2}^{j}\>,
\label{4.2}
\eea
where $D_{i,j}^{(n)}$ are coefficients obtained by inverting the expansion 

\bea
|\psi_n^{i}\> = 
\sum_{j=1}^{4}C_{i,j}^{(n)}|\phi_n^{j}\>. 
\label{3.4.1}
\eea
Applying the spectral 
decomposition of the propagator to (\ref{4.2}) we obtain $\psi(t)$. 
Now that we know the explicit form of the time-dependent state, we are in the position to compute all quantities related to it. Since we are interested in the way in which the bound fermion state entangles with the external degrees 
of freedom, we take a pure state density operator 
$\rho = | \psi(t) \rangle \langle \psi(t) |$ of the entire system
and compute  purity $P$ and entropy $S$ of the Dirac oscillator subsystem:

\bea
P(t)&=& \rm{Tr}_{N,\sigma} \left( \left( \rm{Tr}_{\tau} \rho(t) \right)^2 \right)\nonumber \\
S(t)&=& - \rm{Tr}_{N,\sigma} \left( \rm{Tr}_{\tau} \rho(t) \rm{Log}\left( \rm{Tr}_{\tau} \rho(t) \right) \right),\nonumber \\
\label{4.3}
\eea
where $\rm{Tr}_{N,\sigma}$ is the trace with respect to oscillator and 
$*$-spin degrees of freedom, while $\rm{Tr}_{\tau}$ is the trace 
with respect to isospin. Results for (\ref{4.3}) are depicted in 
figure (\ref{fig2}) as 
functions of parameters $\alpha, \gamma, m$ and time.

One finds two regimes of maximal entanglement as expected. The first one 
is naturally related to the possibility of tuning the external field 
to the frequency of the oscillator, \ie $\alpha \sim 1$. The second regime appears when 
the strength of the coupling is comparable with the rest mass of the 
Dirac particle, \ie $\gamma \sim m$. This shows that the structure of the external field is 
capable of probing the region where particles can be created. It is well known that under these circumstances the external field creates particles and the quantization of the Dirac field becomes important \cite{fw}. At the level of classical fields, the results show that this particle creation is accompanied by a surge in the entanglement between the Dirac oscillator and the external field. It is not surprising, that purity and entropy yield quite similar information. We display both, as entropy is more popular but purity can be obtained analytically. We refer to \cite{varga}, \cite{varga2} for a discussion of the difference of the two quantities.

\section{Conclusions and outlook}

By considering a bound fermion under the influence of a non-local external field, we have constructed a generalization of the Dirac
oscillator in which interactions represent the influence of extended particles in the Yukawa formalism or the non-local gauge field theory. The relativistic model has been shown to be solvable in one, two and three dimensions by constructing the corresponding invariants in a specific frame of reference. By covariance, the solutions in arbitrary inertial frames can be reached by Lorentz transformations.
We gave spectra and eigenstates explicitly for several cases according to different choices of parameters. These include explicit expressions for positive and negative energy (particle and antiparticle) states as well as corrections to the rest mass of the bound fermion. An interpretation of entanglement of the field with positive and negative energy states as a particle creation effect has been possible.

It is worth to mention that our solvable problem has been formulated in a way which allows an immediate identification with quantum optics. The remarkable analogy between the Dirac oscillator and the Jaynes-Cummings hamiltonian \cite{bunch}, \cite{knight} has been pointed out before \cite{delgado}, \cite{solano}, \cite{delgado2} with the aim of producing such a system in a quantum optical experiment. Our work may extend the analogy further by identifying our hamiltonians with a Jaynes-Cummings model of two atoms. The $*-$spin and isospin of the field correspond to the operators describing two atoms of two levels, while the harmonic oscillator operators $a, a^{\dagger}$ correspond to one mode of the electromagnetic field. An experimental realization for this system in the context of quantum optics can be proposed as well. In \cite{nosotros}, we have given a dynamical study in this context. Specifically, we may consider the following setup for the case studied in section 5.1, where the parameters are quite general. We require two atoms of different species trapped in an electromagnetic cavity. Both atoms must be nearly in resonance with only one mode of the eletromagnetic field. Therefore, a sufficiently large spacing between adjacent modes is needed. The coupling constant $\alpha$ defined in relation with $\omega$ (fixed as unity), can be adjusted by placing the atoms in regions with different field intensties. Moreover, a large distance between the atoms is needed in order to ignore direct interaction terms. The case 5.2 is a particular restriction for which the atoms are of the same kind. Finally, the evolution of entanglement studied in section 6 can be realized by preparing the initial state in (\ref{4.2}). In principle, one could prepare it by trapping one atom in the cavity and measuring the energy of the total system (this corresponds to a dressed state of the Jaynes-Cummings model with one atom). After this is achieved, the second atom can be introduced in the trap. This setup thus emulates our model and makes it experimentally accessible.

\ack

T.H. Seligman and E. Sadurn\'i dedicate this work to the memory of their teacher and friend Marcos Moshinsky. We acknowledge support under the Grants PAPIIT IN112507 and CONACyT 57334

\end{document}